\documentclass[journal]{IEEEtran}
\usepackage{amsmath,amsfonts}
\usepackage{algorithmic}
\usepackage{algorithm}
\usepackage{array}
\usepackage[caption=false]{subfig}
\usepackage{textcomp}
\usepackage{stfloats}
\usepackage{url}
\usepackage{verbatim}
\usepackage{graphicx}
\usepackage{cite}
\usepackage{bm}
\usepackage{booktabs}
\usepackage{hyperref}
\usepackage{multirow, multicol}
\usepackage{makecell}
\usepackage{amssymb}
\usepackage{bbding}
\usepackage{enumitem}
\begin{document}

\title{Explicit Estimation of Magnitude and Phase Spectra in Parallel for High-Quality Speech Enhancement}

\author{Ye-Xin Lu, Yang Ai,~\IEEEmembership{Member,~IEEE}, Zhen-Hua Ling,~\IEEEmembership{Senior Member,~IEEE}
\thanks{The authors are with the National Engineering Research Center of Speech and Language Information Processing, University of Science and Technology of China, Hefei 230027, China (e-mail: yxlu0102@mail.ustc.edu.cn; yangai@ustc.edu.cn; zhling@ustc.edu.cn).
}
\thanks{This work is the extended version of our conference paper \cite{lu2023mp} published at Interspeech 2023.}}



\maketitle

\begin{abstract}
Phase information has a significant impact on speech perceptual quality and intelligibility.
However, existing speech enhancement methods encounter limitations in explicit phase estimation due to the non-structural nature and wrapping characteristics of the phase, leading to a bottleneck in enhanced speech quality.
To overcome the above issue, in this paper, we proposed MP-SENet, a novel Speech Enhancement Network that explicitly enhances Magnitude and Phase spectra in parallel.
The proposed MP-SENet comprises a Transformer-embedded encoder-decoder architecture.
The encoder aims to encode the input distorted magnitude and phase spectra into time-frequency representations, which are further fed into time-frequency Transformers for alternatively capturing time and frequency dependencies.
The decoder comprises a magnitude mask decoder and a phase decoder, directly enhancing magnitude and wrapped phase spectra by incorporating a magnitude masking architecture and a phase parallel estimation architecture, respectively. 
Multi-level loss functions explicitly defined on the magnitude spectra, wrapped phase spectra, and short-time complex spectra are adopted to jointly train the MP-SENet model. 
A metric discriminator is further employed to compensate for the incomplete correlation between these losses and human auditory perception.
Experimental results demonstrate that our proposed MP-SENet achieves state-of-the-art performance across multiple speech enhancement tasks, including speech denoising, dereverberation, and bandwidth extension.
Compared to existing phase-aware speech enhancement methods, it further mitigates the compensation effect between the magnitude and phase by explicit phase estimation, elevating the perceptual quality of enhanced speech.
Remarkably, for the speech denoising task, the proposed MP-SENet yields a PESQ of 3.60 on the VoiceBank+DEMAND dataset and 3.62 on the DNS challenge dataset.
\end{abstract}


\section{Introduction}
\IEEEPARstart{I}{n} real-life scenarios, speech signals captured by devices are inevitably distorted by intrusive noises and their self-reflections from the surfaces of surrounding objects. For the bandwidth-limited scenario, when these captured signals are transmitted between narrowband communication devices, only their low-frequency components are typically preserved. 
These issues immensely degrade the performances of automatic speech recognition, hearing aids, and telecommunication systems \cite{weninger2015speech, desjardins2014effect}. 
In this regard, speech enhancement (SE) aims to improve the intelligibility and overall perceptual quality of distorted speech signals.
With the development of deep learning and deep neural network (DNN), numerous DNN-based SE methods have been proposed to specifically address one or more of these issues, with the aim of recovering the original speech signals from the distorted ones. 

For speech signal processing, the phase information determines the temporal and spectral characteristics of the speech signal, exerting a substantial impact on the naturalness and intelligibility of the reconstructed speech.
In convolutional SE methods, restoring phase information has long been underestimated, with distorted phase being used as an optimal estimator of the original phase  \cite{wang1982unimportance}.
However, recent studies demonstrated the noteworthy contribution of precise knowledge of the phase spectrum to speech quality \cite{paliwal2011importance}.
The phase information ensures coherence across time frames and between adjacent frequency bins in the time-frequency (TF) domain.
Accurately preserving it during enhancement helps maintain the temporal characteristics and harmonic structures of the original speech.

Despite the significance of phase reconstruction in SE tasks, existing SE methods still face challenges in precisely modeling and optimizing the phase due to its non-structural nature and wrapping characteristics.
Time-domain SE methods \cite{pascual2017segan, kuleshov2017audio, defossez2020real, kim2021se, wang2021towards} utilized neural networks to learn the mapping from distorted waveforms to original ones, implicitly restored both the magnitude and phase information without explicit modeling.
For the TF-domain phase-aware SE methods, some focused on enhancing the short-time complex spectrum, which implicitly recovered both magnitude and phase within the complex domain \cite{choi2018phase, hao2021fullsubnet, kothapally2022skipconvgan, zhao2022frcrn, dang2022dpt, yin2022tridentse, wang2020complex, wang2023tf}.
Recently, several advancements in the form of multi-stage decoupling-style methods have been proposed, which first explicitly enhanced the magnitude spectrum and subsequently conducted complex spectrum refinement to implicitly enhance the phase \cite{li2021two, li2022taylor, yu2022dual, abdulatif2022cmgan}.
While both time-domain and TF-domain methods can partially recover phase information through modeling and optimizing waveforms or complex spectra, Wang \emph{et al.} \cite{wang2021compensation} pointed out that optimizing loss functions defined solely in the time or complex domain would result in implicit compensation issue between magnitude and phase.
Although introducing additional magnitude-domain loss can effectively alleviate this issue \cite{wang2021compensation}, the lack of explicit phase optimization would still result in magnitude compensating for inaccurate and discontinuous phase prediction, undermining the harmonic structures of the magnitude and impacting the enhanced speech quality.
Therefore, explicit phase modeling and optimization are particularly important for SE models to produce precise magnitude and phase with slighter compensation effect, thereby further elevating the perceptual quality and intelligibility of enhanced speech.

To this end, we propose MP-SENet, a TF-domain monaural SE model that performs parallel estimations of magnitude and phase spectra.
The MP-SENet adopts an encoder-decoder architecture, connecting the encoder and decoder with Transformers \cite{vaswani2017attention} across both time and frequency dimensions, enabling the capture of alternating time and frequency dependencies.
The encoder encodes the input distorted magnitude and wrapped phase spectra derived from short-time Fourier transform (STFT) into TF-domain representations for subsequent decoding. The parallel magnitude mask decoder and phase decoder estimate the original magnitude and phase spectra, respectively, and the enhanced speech waveforms are finally reconstructed using inverse STFT (iSTFT).
To realize explicit phase enhancement, we follow our previous work \cite{ai2022neural} to utilize the parallel estimation architecture and anti-wrapping loss to explicitly model and optimize the phase spectra, respectively.
Additionally, we utilize magnitude loss, complex spectral loss, and STFT consistency loss to train the MP-SENet model effectively.
Furthermore, to address the incomplete correlation between these objective losses and human auditory perception, we incorporate a metric discriminator \cite{fu2019metricgan, fu2021metricgan+} to help enhance the speech perceptual quality.
Experimental results demonstrate that MP-SENet achieves state-of-the-art (SOTA) performance across speech denoising, dereverberation, and bandwidth extension (BWE) tasks. 
It is noteworthy that compared to previous methods with implicit phase optimization, our MP-SENet further alleviates the compensation effect between magnitude and phase by explicitly modeling and optimizing the phase, resulting in more accurate magnitude-phase prediction and harmonic restoration.

The main contribution of this work lies in proposing a comprehensive solution for phase-aware speech enhancement, which involves explicit modeling and optimization of both magnitude and phase. 
This enables the model to predict more accurate magnitude and phase with slighter compensation, thereby elevating the quality of enhanced speech to a new level.
To the best of our knowledge, we are the first to achieve direct enhancement of the phase.
Additionally, our proposed MP-SENet exhibits the ability to handle three distinct SE tasks, i.e., speech denoising, dereverberation, and BWE with a unified framework, showcasing its generalization capability across different SE tasks.

The rest of this paper is organized as follows.
Section~\ref{sec: II} briefly reviews several phase-aware DNN-based SE methods and the relevant backgrounds of three aforementioned SE tasks.
In Section~\ref{sec: III}, we give details of our proposed MP-SENet framework.
The experimental setup is presented in Section~\ref{sec: IV}, while Section~\ref{sec: V} gives the results and analysis. 
Finally, we give conclusions in Section~\ref{sec: VI}.

\section{Related Work}
\label{sec: II}
\subsection{Phase-aware SE Methods}
\label{sec: II.A}
In traditional SE methods, only the magnitude spectrum was enhanced and then combined with the distorted phase spectrum to reconstruct the enhanced speech \cite{xu2014regression, ai2019dnn}.
Upon recognizing the importance of phase reconstruction in improving speech perceptual quality, plenty of phase-aware SE methods have been proposed to recover phase information from distorted speech or spectral features.
In this paper, we primarily focus on the TF-domain phase-aware SE methods and broadly classify them into three categories for discussion.

\begin{itemize}[left=0em]
\item{The first category of TF-domain phase-aware SE methods enhances the phase within the complex spectrum and can be further divided into complex spectral masking-based and complex spectral mapping-based methods.
Complex spectral masking \cite{choi2018phase, hao2021fullsubnet, zhao2022frcrn, kothapally2022skipconvgan, dang2022dpt, yin2022tridentse} is a commonly used method that predicts the complex-valued mask applied to the real and imaginary parts of the distorted short-time complex spectrum to mask the distorted components.
While in recent studies \cite{wang2020complex, wang2023tf}, it seems that complex spectral mapping, which directly maps the distorted complex spectrum to the clean one, tends to have better performance than spectral masking, especially in low signal-to-noise ratio (SNR) conditions.
For both complex spectral masking and mapping, the phase information was implicitly restored when optimizing the complex spectrum.}

\item{The second category is the decouple-style SE methods \cite{li2021two, li2022taylor, abdulatif2022cmgan}, which decomposes the complex spectrum estimation problem into two sub-stages, including magnitude estimation and complex spectral refinement, respectively.
Specifically, in the first stage, the distorted magnitude spectrum undergoes initial enhancement, followed by combining with the distorted phase spectrum to obtain a coarse complex spectrum estimation.
In the second stage,  an auxiliary network predicts the residual complex spectrum which is then added to the preliminary enhanced complex spectrum to obtain a more precise complex spectrum estimation.
This category of methods aims to compensate for the distorted phase by the complex spectral refinement, thereby restoring the intact phase information.
}

\item{Both of the above types implicitly model the phase within the complex spectrum, while the last category of phase-aware SE methods incorporates a dual-stream architecture to explicitly model magnitude and phase \cite{yin2020phasen}.
Similar to the decouple-style methods, the distorted magnitude is firstly enhanced through magnitude masking in the magnitude stream, and the phase stream outputs a complex-valued feature map with two channels, corresponding to the real and the imaginary parts.
Differently, the magnitude of the complex feature map is normalized to 1, and consequently, the complex feature map only contains the phase information.
In this way, the phase spectrum can be explicitly modeled by the complex spectrum features.}
\end{itemize}

In summary, existing TF-domain phase-aware SE methods implicitly or explicitly model the phase by predicting the complex spectrum. However, they still optimize the phase implicitly in the complex spectral domain, leading to a compensation effect \cite{wang2021compensation} between magnitude and phase. 
Even with the introduction of magnitude-domain loss, phase optimization remains challenging, resulting in magnitude compensation for inaccurate and discontinuous phase prediction, thereby affecting its harmonic structure and further impacting speech perceptual quality.
Thus, there is scope for improvement in terms of speech perceptual quality and intelligibility by undertaking both explicit phase modeling and optimization.

\subsection{Speech Enhancement Tasks}
In this paper, we mainly focus on three SE tasks, including speech denoising, dereverberation, and BWE.
In this sub-section, we will sequentially introduce the backgrounds and relevant methods for these three tasks.

\subsubsection{Speech denoising}
Speech denoising aims to remove the additive noise signal from the noisy speech signal. 
Classic signal-processing-based speech denoising methods include spectral subtraction \cite{berouti1979enhancement}, Wiener filtering \cite{lim1978all}, statistical model-based methods \cite{ephraim1992statistical}, subspace algorithms \cite{dendrinos1991speech, ephraim1995signal}, etc.
These traditional methods often required prior information and faced challenges in dealing with non-stationary noise.
With the renaissance of deep learning, DNN-based denoising methods demonstrate strong non-stationary noise suppression capabilities and achieve better SE performance.
Existing DNN-based denoising methods can be classified into two categories, i.e., time-domain methods and TF-domain methods.

Time-domain denoising methods aim to predict clean speech waveforms directly from noisy ones without any frequency transformation.
One of the most notable methods in this category is SEGAN \cite{pascual2017segan}, which employs generative adversarial networks (GANs) to learn the mapping from noisy speech waveforms to clean ones at the waveform level.
While this category of methods successfully avoids the phase estimation problem encountered in TF-domain methods, they sacrifice the ability to capture speech phonetics in the frequency domain.
Consequently, this trade-off can lead to artifacts in enhanced speech waveforms.
As a result, the performance of time-domain denoising methods is generally considered inferior to that of TF-domain methods.

TF-domain denoising methods are predominantly based on spectral mapping or masking.
Mapping-based methods map the spectral representations of degraded speech to those of clean speech, including magnitude mapping with noisy phase retained \cite{xu2014regression, ai2019dnn}, complex spectral mapping \cite{wang2020complex, wang2023tf}, and other phase-contained spectral mappings \cite{liu2023mask}.
Masking-based methods initially enhance the magnitude spectrum using the ideal binary mask (IBM) \cite{hu2001speech}, ideal ratio mask (IRM) \cite{srinivasan2006binary, narayanan2013ideal}, and spectral magnitude mask (SMM) \cite{wang2014training} to selectively attenuate the noise components based on the mask values.
Recognizing the significance of phase information, phase-sensitive mask (PSM) \cite{erdogan2015phase} and complex ideal ratio mask (cIRM) \cite{williamson2015complex} were proposed based on SMM and IRM, respectively.
PSM remains a real-valued mask with additional phase estimation, allowing the estimated magnitude to compensate for the noisy phase, while cIRM is a complex-valued mask used to directly enhance the complex spectrum.
In addition, the decoupling-style methods and dual-stream magnitude and phase estimation methods we mentioned in section~\ref{sec: II.A} also belong to this category of methods.

\subsubsection{Speech dereverberation}
Speech dereverberation aims to remove the convolutive effects or reflections caused by the surfaces of surrounding objects, which result in smearing effects across the time and frequency domains of the original speech waveforms.
The quantity and intensity of these reflections are influenced by factors such as room size, surface properties, and microphone distance. 
These reflections can be effectively modeled using the room impulse response (RIR) filter.

In the early research, signal processing techniques were employed to eliminate the convolutive effects caused by reverberation via statistical modeling of the RIRs. 
For instance, the weighted prediction error (WPE) algorithm \cite{nakatani2010speech} employs long-term linear prediction to estimate the impact of RIRs on reverberant speech waveforms, which assumes that the early reflections of RIRs, along with the direct path, are crucial for improved recognition, while late reflections are considered diffused and detrimental to speech intelligibility.

DNN-based speech dereverberation methods are generally similar to the approaches used in denoising tasks, and there are models capable of simultaneously handling both denoising and dereverberation tasks.
Since reverberation manifests as smearing effects in the TF-domain spectrum, IBM \cite{roman2011intelligibility} and IRM \cite{li2017ideal} based methods predict masks applied to the reverberant magnitude spectra, combined with the unprocessed reverberant phase spectra, to reconstruct the anechoic speech waveforms.
Following the speech denoising task, cIRM is further employed to implicitly reconstruct the phase \cite{kothapally2022skipconvgan}.
Furthermore, some mapping-based methods have also achieved promising dereverberation performance \cite{ernst2018speech, kothapally2024monaural}.

\subsubsection{Speech bandwidth extension}
Speech BWE aims to predict the high-frequency components from the narrowband speech signal to obtain a wideband one, thereby extending the frequency bandwidth and enhancing the quality and intelligibility of the speech signal.
Therefore, it can also be regarded as an SE task.
Traditional speech BWE methods utilized signal processing techniques to predict high-frequency residual signals and spectral envelopes.
However, these traditional methods suffered from bottlenecks in terms of model capabilities, leading to the production of overly smoothed spectral parameters \cite{ling2015deep}, whereas recent DNN-based methods have demonstrated superior performance.

DNN-based speech BWE methods can also be classified into time-domain methods, TF-domain methods, and hybrid-domain methods.
Time-domain BWE methods \cite{kuleshov2017audio, wang2021towards} directly predict the high-resolution speech waveforms from low-resolution ones.
For TF-domain BWE methods, the recovery of the high-frequency phase is the major challenge.
Early studies directly replicate \cite{abel2018simple} or mirror inverse \cite{li2015deep} the narrowband phase spectrum to obtain the upper-band phase spectrum.
Vocoder-based methods \cite{liu2022neural} adopt a neural vocoder to restore the wideband speech waveform from the extended mel-spectrogram, and further replace the low-frequency components of the vocoder's output with the original low-frequency speech signal.
Similar to speech denoising and dereverberation, decouple-style methods \cite{abdulatif2022cmgan} employ an additional complex spectral refinement to compensate for the narrowband phase, implicitly recovering the high-frequency phase.
Additionally, the hybrid-domain methods \cite{lim2018time} combine both time-domain and TF-domain information with a dual-branch architecture.
One branch predicts the wideband magnitude spectrum from the narrowband one, and the other predicts the wideband waveform from the input narrowband one.
Finally, the prediction of the wideband phase spectrum is derived from the wideband waveform output of the time-domain branch. 

\begin{figure*}[t!]
  \centering
  \includegraphics[width=\textwidth]{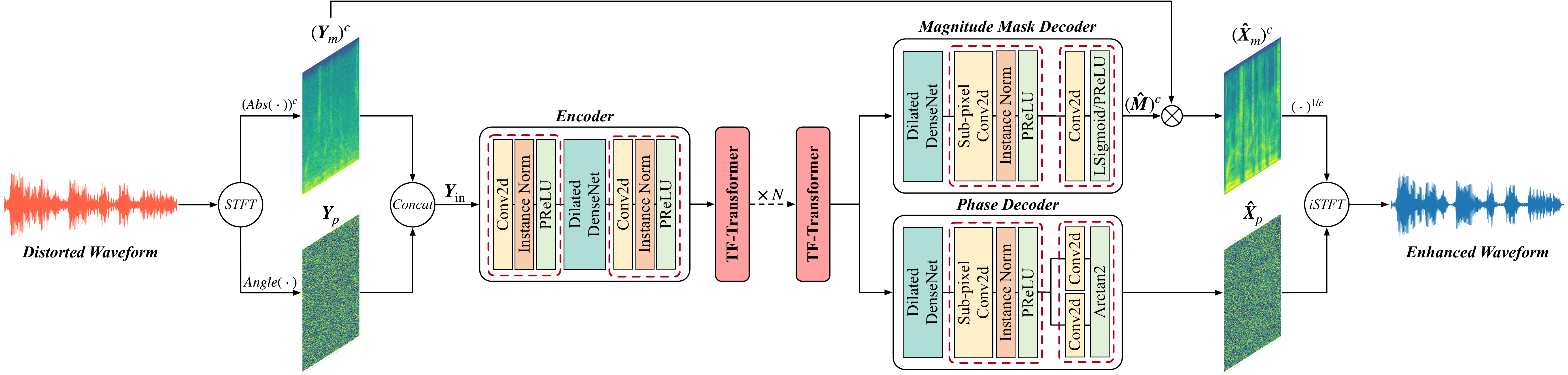}
  \caption{Overall structure of the proposed MP-SENet for three SE tasks, with the denoising task used as an illustration. The ``$Abs( \cdot )$'' and ``$Angle( \cdot )$'' represent the magnitude and phase calculation functions, respectively. The ``$( \cdot )^c$" and ``$( \cdot )^{1/c}$" denote the magnitude compression and decompression operations, respectively. The ``\emph{Concat}'' denotes the spectral concatenation operation and the ``\emph{Arctan2}'' denotes the two-argument arc-tangent function.}
  \label{fig: model}
\end{figure*}

\section{Methodology}
\label{sec: III}
\subsection{Model Structure}
\subsubsection{Overview}
An overview of the model structure of the proposed MP-SENet is illustrated in Fig.~\ref{fig: model}.
For all the speech denoising, dereverberation, and BWE tasks, they share the same architecture.
Generally, the MP-SENet employs an encoder-decoder architecture to enhance the distorted speech waveform $\bm{y} \in \mathbb{R}^L$ and recover the original speech waveform $\bm{x} \in \mathbb{R}^L$ in the TF domain, where $L$ represents the length of the waveform.
To begin, we apply STFT on $\bm{y}$ to extract the short-time complex spectrum $\bm{Y} \in \mathbb{C}^{T \times F}$, and subsequently calculate the magnitude spectrum $\bm{Y}_m \in \mathbb{R}^{T\times F}$ and the wrapped phase spectrum $\bm{Y}_p \in \mathbb{R}^{T\times F}$, where $\bm{Y} = \bm{Y}_m \cdot e^{j\bm{Y}_p}$, and $T$ and $F$ denote the total number of frames and frequency bins, respectively.
Before feeding them to the encoder, we apply the power-law compression \cite{wisdom2019differentiable} on $\bm{Y}_m$ and get the compressed magnitude spectrum $(\bm{Y}_m)^c \in \mathbb{R}^{T\times F}$, where $c$ is the compression factor.
The input feature $\bm{Y}_\mathrm{in} \in \mathbb{R}^{T \times F \times 2}$ of MP-SENet is the concatenation of the $(\bm{Y}_m)^c$ and $\bm{Y}_p$.
The encoder encodes $\bm{Y}_\mathrm{in}$ into a compressed TF-domain representation, and subsequently, the TF-domain representation is processed by $N$ time-frequency transformer (TF-Transformer) blocks to alternately capture time and frequency dependencies.
In the final stages, the parallel magnitude mask decoder and phase decoder predict the compressed magnitude spectrum $(\bm{\hat{X}}_m)^c \in \mathbb{R}^{T \times F}$ and wrapped phase spectrum $\bm{\hat{X}}_p \in \mathbb{R}^{T \times F}$ from the TF-domain representation, respectively.
Subsequently, the enhanced waveform $\bm{\hat{x}}$ is reconstructed through iSTFT.
Further insights into the encoder, TF-Transformer block, magnitude mask decoder, and phase decoder are elaborated below.
\begin{figure}[t!]
  \centering
  \includegraphics[width=\linewidth]{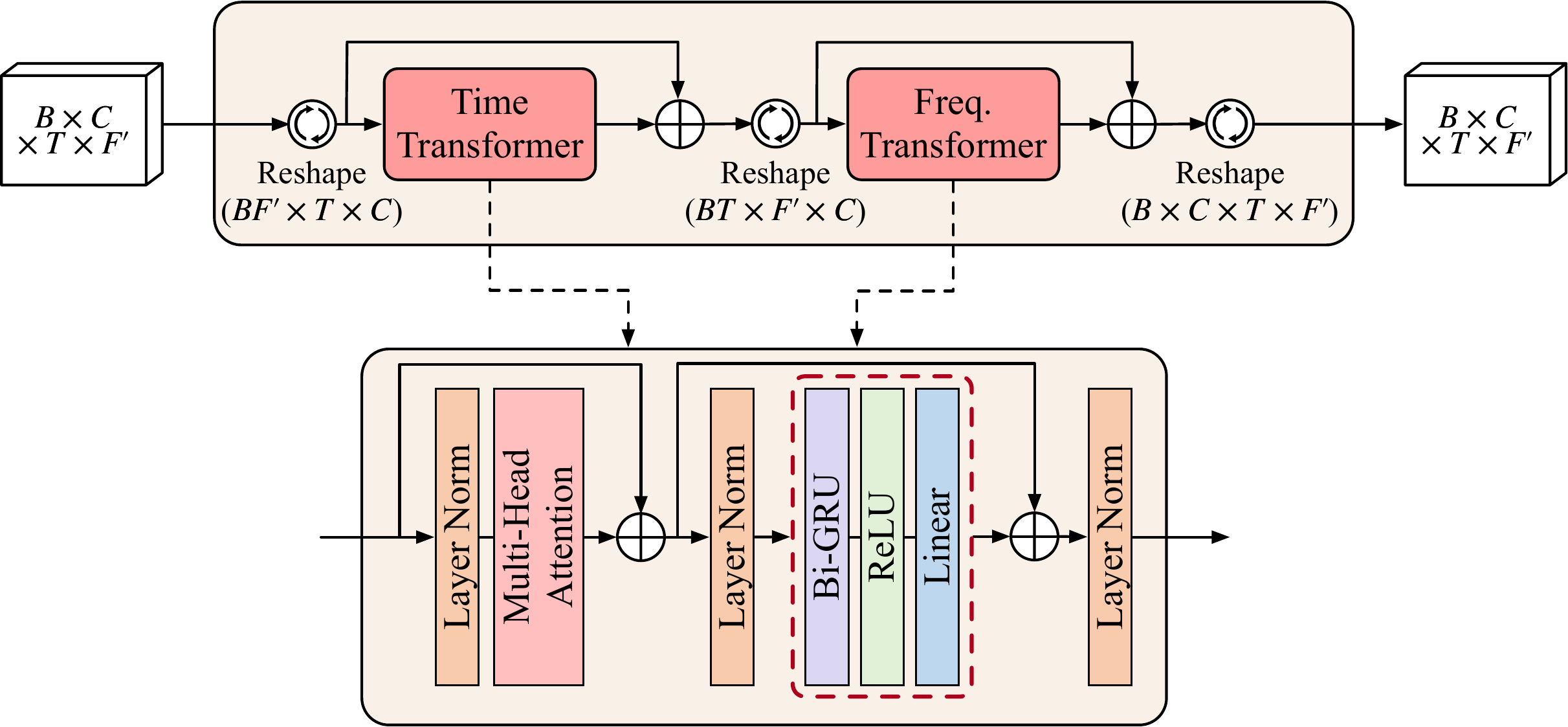}
  \caption{The diagram of the TF-Transformer block used in Fig~\ref{fig: model}, where $B$, $C$, $T$, and $F'$ represent the batch size, the number of channels, the number of frames, and the number of frequency bins, respectively. The TF-Transfomer block is a cascade of a Time-Transformer and a Freq.-Transformer, both of which share the same architecture with inputs of different shapes.}
  \label{fig: TF-Transformer}
\end{figure}
\subsubsection{Encoder}
As illustrated in Fig.~\ref{fig: model}, the encoder transforms the input feature $\bm{Y}_\mathrm{in}$ into a TF-domain representation with $C$ channels, $T$ temporal dimensions, and $F'=F/2$ frequency dimensions.
Specifically, the encoder comprises a cascade of a convolutional block, a dilated DenseNet \cite{pandey2020densely}, and another convolutional block.
Each convolutional block comprises a 2D convolutional layer, an instance normalization \cite{ulyanov2016instance}, and a parametric rectified linear unit (PReLU) activation \cite{he2015delving}.
The first convolutional block increases the channel numbers of the TF-domain representation to $C$, and the second convolutional block halves the feature dimension $F$ to $F'$ to reduce the subsequent computational complexity in the TF-Transformer blocks.
The dilated DenseNet utilizes four convolutional layers with dilation sizes of 1, 2, 4, and 8 to extend the receptive field along the time axis, facilitating context aggregation at different resolutions.
Additionally, it incorporates dense connections at each convolutional layer, connecting to all previous layers.
This not only helps prevent the vanishing gradient problem but also enhances the parameter efficiency of the network.

\subsubsection{TF-Transformer}
The dual-path attention-based structure \cite{wang2021tstnn, fu2022uformer, abdulatif2022cmgan} has demonstrated remarkable performance in SE tasks, primarily due to its ability to capture both time and frequency dependencies.
Draw inspiration from the two-stage Conformers in CMGAN \cite{abdulatif2022cmgan}, we proposed the TF-Transformer block, which can sequentially capture long-range correlations exhibited in the temporal and spectral domains.
As illustrated in Fig.~\ref{fig: TF-Transformer}, the TF-Transformer block takes an intermediate TF-domain representation as input, with a shape of $B\times C \times T \times F'$, where $B$ denotes the batch size.
Initially, this representation is reshaped to $BF'\times T \times C$ and passed through the Time-Transformer layer to capture temporal dependencies.
Subsequently, it is reshaped to $BT\times F' \times C$ and fed into the Frequency-Transformer layer to capture frequency dependencies.
Finally, the processed TF-domain representation is reshaped back to $B\times C \times T \times F'$ as the output of the TF-Transformer block.

For each TF-Transformer block, both the Time-Transformer layer and the Frequency-Transformer layer share the same architecture, which is a gated recurrent unit (GRU) based Transformer \cite{wang2021tstnn} without positional encoding.
The GRU-based transformer consists of a multi-head self-attention (MHSA) \cite{vaswani2017attention} and a GRU-based position-wise feed-forward network (FFN), augmented with layer normalizations \cite{ba2016layer} and residual connections.
The MHSA with $M$ heads enables the model to collectively focus on the global information from different representation subspaces at various positions. 
The GRU-based position-wise FFN comprises a bidirectional GRU (Bi-GRU) layer, a rectified linear unit (ReLU) activation \cite{glorot2011deep}, and a linear layer to learn the local information \cite{sperber2018self}.

\subsubsection{Magnitude mask decoder}
As illustrated in Fig~\ref{fig: model}, the magnitude mask decoder predicts a compressed magnitude mask $(\bm{\hat{M}})^c \in \mathbb{R}^{T \times F}$ from the TF-domain representation processed by the TF-Transformer blocks and multiplies it with the compressed distorted magnitude spectrum $(\bm{Y}_m)^c$ to obtain the enhanced compressed magnitude spectrum $(\bm{\hat{X}}_m)^c$.

The magnitude mask decoder comprises a dilated DenseNet, a deconvolutional block, and a mask estimation architecture.
The deconvolutional block is composed of a 2D sub-pixel convolutional layer \cite{shi2016real}, an instance normalization, and a PReLU activation.
The sub-pixel convolutional layer is used to upsample the frequency dimensions of the TF-domain representation back to $F$ while avoiding checkerboard artifacts compared to transposed convolution.
The mask estimation architecture comprises a 2D convolutional layer and an activation function.
The convolutional layer is used to reduce the output channel numbers of the deconvolutional block to 1, and the activation function is for mask estimation.
For speech denoising or dereverberation, we use the learnable sigmoid (LSigmoid) function\cite{fu2021metricgan+} to estimate a boundary magnitude mask:
\begin{equation}
	{\rm LSigmoid}(t) = \frac{\beta}{1+e^{1-\alpha t}},
\end{equation}
where $\beta$ is set to 2.0, and $\alpha \in \mathbb{R}^F$ is a trainable frequency-related parameter, which enables the model to adaptively adjust the shape of the activation function across frequency bands.
For speech BWE, we use the PReLU activation to predict an unbounded high-frequency magnitude mask.

\subsubsection{Phase decoder}
Considering the phase wrapping problem, we design the phase decoder to directly predict the enhanced wrapped phase spectrum $\bm{\hat{X}}_p$ from the TF-domain representation.
The phase decoder and the magnitude mask decoder share a similar architecture.
To address the challenges posed by the non-structural nature and wrapping characteristics of the phase, we follow our previous work \cite{ai2022neural} and incorporate the parallel estimation architecture after a dilated DenseNet and a deconvolutional block in the phase decoder.
The parallel estimation architecture first adopts two parallel 2D convolutional layers to output the pseudo-real part component and pseudo-imaginary part component and then activates these two components to obtain the enhanced wrapped phase spectrum $\bm{\hat{X}}_p$ using the two-argument arc-tangent (Arctan2) function.

\subsection{Training Criteria}
\subsubsection{Spectrum-based losses}
We first define loss functions on the magnitude spectra, phase spectra, and short-time complex spectra to jointly train the proposed MP-SENet in the spectral domain. 
These loss functions are described as follows.

\begin{itemize}[left=0em]
\item {}{\textbf{Magnitude Spectrum Loss}}: 
The magnitude spectrum loss is defined as the mean square error (MSE) loss between the original compressed magnitude spectrum $(\bm{X}_m)^c \in \mathbb{R}^{T \times F}$ and the enhanced compressed magnitude spectrum $(\bm{\hat{X}}_m)^c$ to explicitly optimize the magnitude spectrum:
\begin{equation}
	\mathcal{L}_{\mathrm{Mag.}} = \mathbb{E}_{(\bm{X}_m, \bm{\hat{X}}_m)} \bigg[\big\Vert (\bm{X}_m)^c - (\bm{\hat{X}}_m)^c \big\Vert_{\mathrm{F}}^2\bigg],
\end{equation}
where $\Vert\cdot\Vert_{\mathrm{F}}$ represents the Frobenius norm.

\item {}{\textbf{Phase Spectrum Loss}}:
Considering the phase wrapping property, the absolute distance between the original phase spectra $\bm{X}_p \in \mathbb{R}^{T \times F}$ and the enhanced phase spectra $\bm{\hat{X}}_p$ may not accurately reflect their actual distance, revealing the inappropriateness of conventional phase losses (e.g., absolute $L^p$ distance) for phase optimization.
Thus, to explicitly optimize the phase spectrum, we utilize the anti-wrapping function we proposed in \cite{ai2022neural} to prevent error expansion caused by phase wrapping, which is defined as:
\begin{equation}
    f_{\rm AW}(t) = \big\vert t - 2\pi \cdot {\rm round} \big(\frac{t}{2\pi}\big) \big\vert, t \in \mathbb{R}.
\end{equation}

On the basis of the anti-wrapping function, we first define the instantaneous phase (IP) loss between the original phase spectrum $\bm{X}_p $ and the enhanced phase spectrum $\bm{\hat{X}}_p$:
\begin{equation}
	\mathcal{L}_{\rm IP} = \mathbb{E}_{(\bm{X}_p, \bm{\hat{X}}_p)} \bigg[\big\Vert f_{\rm AW}(\bm{X}_p - \bm{\hat{X}}_p) \big\Vert_1\bigg].
\end{equation}
Considering the phase continuity along both the time and frequency axis, we calculate the group delay (GD) spectrum and the instantaneous angular frequency (IAF) spectrum of the phase, which correspond to the derivatives of the phase spectrum along the frequency axis and the time axis, respectively.
We further define the GD loss between the anti-wrapped original and enhanced GD spectra as well as the IAF loss between the anti-wrapped original and enhanced IAF spectra:
\begin{equation}
    \mathcal{L}_{\rm GD} = \mathbb{E}_{\Delta_{\rm DF} (\bm{X}_p, \bm{\hat{X}}_p)} \bigg[\big\Vert f_{\rm AW}(\Delta_{\rm DF}(\bm{X}_p - \bm{\hat{X}}_p)) \big\Vert_1\bigg],
\end{equation}
\begin{equation}
    \mathcal{L}_{\rm IAF} = \mathbb{E}_{\Delta_{\rm DT} (\bm{X}_p, \bm{\hat{X}}_p)} \bigg[\big\Vert f_{\rm AW}(\Delta_{\rm DT}(\bm{X}_p - \bm{\hat{X}}_p)) \big\Vert_1\bigg],
\end{equation}
where $\Delta_{DF}$ and $\Delta_{DT}$ represent the differential operator along the frequency axis and temporal axis, respectively.

The final phase spectrum loss is the sum of IP loss, GD loss, and IAF loss:
\begin{equation}
	\mathcal{L}_{\rm Pha.} = \mathcal{L}_{\rm IP} + \mathcal{L}_{\rm GD} + \mathcal{L}_{\rm IAF}.
\end{equation}

\item {}{\textbf{Complex Spectrum Loss}}:
To further optimize the magnitude and phase spectra within the complex domain, we define the complex spectrum loss between the real and imaginary parts of the original complex spectrum $\bm{X} = \bm{X}_m \cdot e^{j\bm{X}_p} \in \mathbb{C}^{T \times F}$ and the enhanced complex spectrum $\bm{\hat{X}} = \bm{\hat{X}}_m \cdot e^{j\bm{\hat{X}}_p} \in \mathbb{C}^{T \times F}$ as follows:
\begin{equation}
    \mathcal{L}_{\mathrm{Com.}} = \mathbb{E}_{(\bm{X}, \bm{\hat{X}})} \bigg[\big\Vert \mathrm{Real}(\bm{X} - \bm{\hat{X}}) \big\Vert_{\mathrm{F}}^2 + \big\Vert \mathrm{Imag}(\bm{X} - \bm{\hat{X}}) \big\Vert_{\mathrm{F}}^2\bigg],
\end{equation}
where $\mathrm{Real}(\cdot)$ and $\mathrm{Imag}(\cdot)$ represent operations for extracting the real and imaginary parts from the complex spectrum, respectively.

\item {}{\textbf{STFT Consistency Loss}}:
The previous losses are directly defined on the magnitude spectrum, phase spectrum, and their complex combination output by the model, without considering the inconsistency introduced by the iSTFT and STFT \cite{le2008explicit, le2010fast}.
Therefore, to ensure consistency between $\bm{\hat{X}}$ and its consistent complex spectrum $\mathrm{STFT}(\mathrm{iSTFT}(\bm{\hat{X}}))$, we define the STFT consistency loss between the real and imaginary parts of them as follows:
\begin{equation}
    \begin{aligned}
         \mathcal{L}_{\rm Con.} = \mathbb{E}_{\bm{\hat{X}}} \bigg[\big\Vert &\mathrm{Real}(\bm{\hat{X}} - \mathrm{STFT}(\mathrm{iSTFT}(\bm{\hat{X}})) \big\Vert_{\mathrm{F}}^2 \\
         + \big\Vert &\mathrm{Imag}(\bm{\hat{X}} - \mathrm{STFT}(\mathrm{iSTFT}(\bm{\hat{X}})) \big\Vert_{\mathrm{F}}^2\bigg]
    \end{aligned}
\end{equation}

\end{itemize}

\subsubsection{MetricGAN-based losses}
The primary goal of speech enhancement is to improve the quality and intelligibility of distorted speech considering human perception \cite{benesty2006speech}.
However, specific objective metrics associated with human perception, such as the perceptual evaluation of speech quality (PESQ) \cite{rix2001perceptual} and short-time objective intelligibility (STOI) \cite{taal2011algorithm}, are non-differentiable.
Thus they cannot be directly used as loss functions.
To overcome this issue, MetricGAN \cite{fu2019metricgan} introduces a metric discriminator as a learned surrogate of the evaluation metrics. 
The metric discriminator iteratively estimates a surrogate loss that approaches the sophisticated metric surface, enabling the SE model to leverage this surrogate to determine the optimization direction toward achieving improved speech perceptual quality.

We here adopt the metric discriminator from \cite{abdulatif2022cmgan} for adversarial training, which uses the PESQ as the target objective metric.
The metric discriminator uses pairs of reference and enhanced speech waveforms as inputs and outputs the corresponding PESQ score of the enhanced speech from their magnitude.
Since the original PESQ score ranges from -0.5 to 4.5, we scale them to the range of 0 to 1 with linear normalization.
With the metric discriminator, we subsequently define the metric loss to guide the MP-SENet model training in generating speech waveforms corresponding to as high PESQ scores as possible.
The discriminator loss and the corresponding generator metric loss are described as follows:
\begin{equation}
	\mathcal{L}_{\rm D} = \mathbb{E}_{\bm{x}} \bigg[\big\Vert D(\bm{x}, \bm{x})-1 \big\Vert_{\mathrm{F}}^2\bigg] + \mathbb{E}_{(\bm{x}, \bm{\hat{x}})} \bigg[\big\Vert D(\bm{x}, \bm{\hat{x}})- Q_{\rm PESQ} \big\Vert_{\mathrm{F}}^2\bigg],
\end{equation}
\begin{equation}
	\mathcal{L}_{\rm Metric} = \mathbb{E}_{(\bm{x}, \bm{\hat{x}})} \bigg[\big\Vert D(\bm{x}, \bm{\hat{x}})-1 \big\Vert_{\mathrm{F}}^2\bigg],
\end{equation}
where $D$ denotes the metric discriminator and $Q_{\rm PESQ} \in [0,1]$ denotes the real scaled PESQ score between $\bm{x}$ and $\bm{\hat{x}}$.

\subsubsection{The final loss for model training}
The final generator loss is the linear combination of the aforementioned spectrum-based losses and the generator metric loss:
\begin{equation}
	\mathcal{L}_{\rm G} = \lambda_1\mathcal{L}_{\rm Mag.} + \lambda_2\mathcal{L}_{\rm Pha.} + \lambda_3\mathcal{L}_{\rm Com.} + \lambda_4\mathcal{L}_{\rm Con.} + \lambda_5\mathcal{L}_{\rm Metric},
\end{equation}
where $\lambda_1$, $\lambda_2$, ..., $\lambda_5$ are hyperparameters. The training criteria of the MP-SENet is to jointly minimize $\mathcal{L}_{\rm G}$ and $\mathcal{L}_{\rm D}$.

\section{Experimental Setup}
\label{sec: IV}
\subsection{Datasets}
\subsubsection{Speech denoising}
We used the publicly available dataset \cite{valentini2016investigating} for the speech denoising experiments, including the VoiceBank+DEMAND dataset and the Deep Noise Suppression (DNS) Challenge 2020 dataset \cite{reddy2020interspeech}.

The VoiceBank+DEMAND dataset includes pairs of clean and noisy audio clips with a sampling rate of 48 kHz.
All the audio clips were resampled to 16 kHz in the experiments.
The clean audio set is selected from the Voice Bank corpus \cite{veaux2013voice}, which consists of 11,572 audio clips from 28 speakers for training and 824 audio clips from 2 unseen speakers for testing.
The clean audio clips are mixed with 10 types of noise (8 types from the DEMAND database \cite{thiemann2013diverse} and 2 artificial types) at SNRs of 0dB, 5dB, 10dB, and 15 dB for the training set and 5 types of unseen noise (i.e., living room, office space, bus, open area cafeteria, and public square) from the DEMAND database at SNRs of 2.5 dB, 7.5dB, 12.5 dB, and 17.5 dB for the test set.

The DNS Challenge 2020 dataset includes 500 hours of clean audio clips from 2150 speakers and over 180 hours of noise audio clips.
All the clean audio clips were split into segments of 10 seconds.
Following the official script \footnote{\href{https://github.com/microsoft/DNS-Challenge}{https://github.com/microsoft/DNS-Challenge}.}, we generated 3,000 hours of noisy audio clips with SNRs from -5dB to 15dB for training.
For model evaluation, the DNS Challenge 2020 dataset provided non-blind test sets that contained data with or without reverberation.
Since we used a dedicated dataset for the dereverberation task, we here only adopted the test set without reverberation.
The test set contained 4,270 audio clips spoken by 20 speakers and corresponding noisy audio clips with SNR levels uniformly sampled between 0 dB and 25 dB.

\subsubsection{Speech dereverberation}
We used the Reverb Challenge dataset \cite{kinoshita2016summary} for the speech dereverberation experiments, exclusively employing its single-channel data for the monaural speech enhancement conducted in this study.
The training set consisted of the clean WSJCAM0 \cite{robinson1995wsjcam0} training set and a multi-condition training set, which was generated from the clean WSJCAM0 training data by convolving these clean utterances with 24 measured RIRs with the reverberation time ($T_{60}$) ranging from 0.2s to 0.8s and adding recorded background noise at a fixed SNR of 20 dB.
The validation set and the test set each consisted of two different parts, namely the `SimData' and the `RealData'.
The `SimData' consists of artificially distorted versions of utterances taken from the WSJCAM0 corpus.
These data were created by convolving the clean WSJCAM0 signals with RIRs measured in three different rooms (small, medium, and large) with $T_{60}$ of approximately 0.25s, 0.5s, and 0.7s, and microphone distance of a near condition (0.5m) and a far condition (2m).
Consistent with the training set, recorded background noise was added to the reverberated data at a fixed SNR of 20 dB.
The `RealData' consists of real reverberant speech recordings from the MC-WSJ-AV dataset \cite{lincoln2005multi}, which were captured in a noisy and reverberant meeting room with $T_{60}$ of approximately 0.7s and microphone distance of 1m and 2.5m.

\subsubsection{Speech bandwidth extension}
We used the VCTK dataset \cite{veaux2017cstr} which contains 44 hours of speech recordings from 108 speakers with a sampling rate of 48 kHz for the speech BWE experiments.
We followed the previous works \cite{kuleshov2017audio, lim2018time, abdulatif2022cmgan} to use the first 100 speakers for training and the remaining 8 speakers for testing.
To generate the wideband-narrowband speech pairs, We first resampled all the audio clips to the sampling rate of 16 kHz as the wideband speech.
We further followed the implementation of AudioUNet \footnote{\href{https://github.com/kuleshov/audio-super-res}{https://github.com/kuleshov/audio-super-res}.} to subsampled the 16 kHz wideband speech into narrowband ones with a subsampling rate, and upsampled them back to 16 kHz using the spline interpolation.
We set the subsampling rate to 2 and 4 in our implementation, indicating the extensions from 8 kHz and 4 kHz to 16 kHz, respectively.

\begin{table*}[htbp!]\scriptsize
  \caption{Experimental results for speech denoising methods evaluated on the VoiceBank+DEMAND dataset. ``-'' denotes the result that is not provided in the original paper.}
  \label{tab: denoising_vb}
  \centering
  \resizebox{0.9\textwidth}{!}{
  \begin{tabular}{lclcccccc}
    \toprule
    Method & Year & Structure & \#Param. & WB-PESQ & CSIG & CBAK & COVL & STOI \\
    \midrule
    Noisy          					   & -     & -                  & -    & 1.97 & 3.35 & 2.44 & 2.63 & 0.91 \\
    \midrule
    SEGAN \cite{pascual2017segan}      & 2017  & Waveform           & 43.2M & 2.16 & 3.48 & 2.94 & 2.80 & 0.92 \\
    DEMUCS \cite{defossez2020real}     & 2021  & Waveform           & 33.5M & 3.07 & 4.31 & 3.40 & 3.63 & 0.95 \\
    SE-Conformer \cite{kim2021se}      & 2021  & Waveform           & -    & 3.13 & 4.45 & 3.55 & 3.82 & 0.95 \\
    MetricGAN \cite{fu2019metricgan}   & 2019  & Magnitude          & -    & 2.86 & 3.99 & 3.18 & 3.42 & -    \\
    MetricGAN+ \cite{fu2021metricgan+} & 2021  & Magnitude          & -    & 3.15 & 4.14 & 3.16 & 3.64 & -    \\
    DPT-FSNet  \cite{dang2022dpt}      & 2021  & Complex            & 0.88M & 3.33 & 4.58 & 3.72 & 4.00 & \textbf{0.96} \\
    TridentSE \cite{yin2022tridentse}  & 2023  & Complex            & 3.03M & 3.47 & 4.70 & 3.81 & 4.10 & \textbf{0.96} \\
    DB-AIAT \cite{yu2022dual}          & 2021  & Magnitude+Complex  & 2.81M & 3.31 & 4.61 & 3.75 & 3.96 & -    \\
    CMGAN \cite{abdulatif2022cmgan}          & 2022  & Magnitude+Complex  & 1.83M & 3.41 & 4.63 & 3.94 & 4.12 & \textbf{0.96} \\
    PHASEN \cite{yin2020phasen}        & 2020  & Magnitude+Phase    & 20.9M & 2.99 & 4.21 & 3.55 & 3.62 & -    \\
    \textbf{MP-SENet} & 2024  & Magnitude+Phase & 2.26M & \textbf{3.60} & \textbf{4.81} & \textbf{3.99} & \textbf{4.34} & \textbf{0.96} \\
    \bottomrule
  \end{tabular}
  }
\end{table*}

\subsection{Model Configuration}
In the training stage, all the audio clips were sliced into 2-second segments.
To extract input features from raw waveforms using STFT, the FFT point number, Hanning window size, and hop size were set to 400, 400 (25 ms), and 100 (6.25 ms), and consequently, the number of frequency bins $F=201$.
For compressing the magnitude spectrum, the compression factor $c$ was set to 0.3.
For training our proposed MP-SENet, we set the batch size $B$, the number of channels $C$, the number of TF-Transformer blocks $N$, and the number of heads in the MHSA $M$ to 4, 64, 4, and 4, respectively.
The hyperparameters of the final generator loss $\lambda_1$, $\lambda_2$, ..., $\lambda_5$ were set to 0.9, 0.3, 0.1, 0.1, and 0.05 with the empirical trial.
All the models were trained until 500k steps using the AdamW optimizer \cite{loshchilov2017decoupled}, with $\beta_1=0.8$, $\beta_2=0.99$, and weight decay of $0.01$.
The learning rate was set initially to 0.0005 and scheduled to decay with a factor of 0.99 every epoch. \footnote{Audio samples of the proposed MP-SENet can be accessed at the official website \href{https://yxlu-0102.github.io/MP-SENet}{https://yxlu-0102.github.io/MP-SENet}.}

\begin{figure*}[!t]
\centering
\subfloat[VoiceBank+DEMAND dataset]{\includegraphics[width=\textwidth]{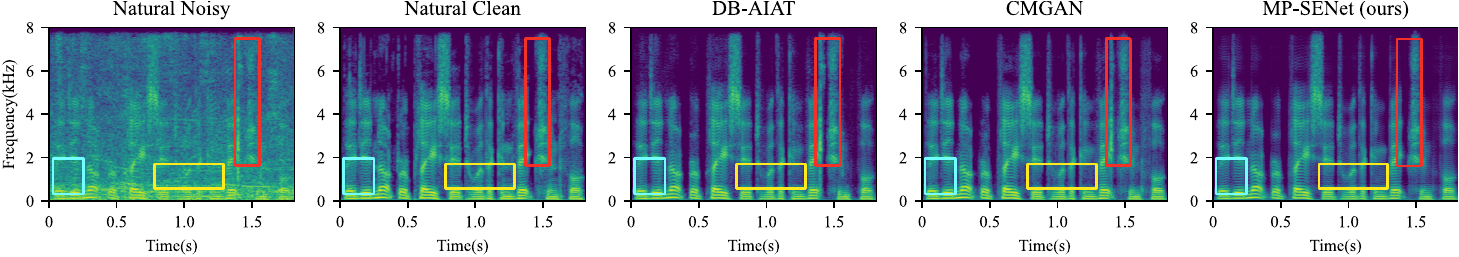}
\label{fig: specs_denoising_vb}}
\hfill
\subfloat[DNS Challenge dataset]{\includegraphics[width=\textwidth]{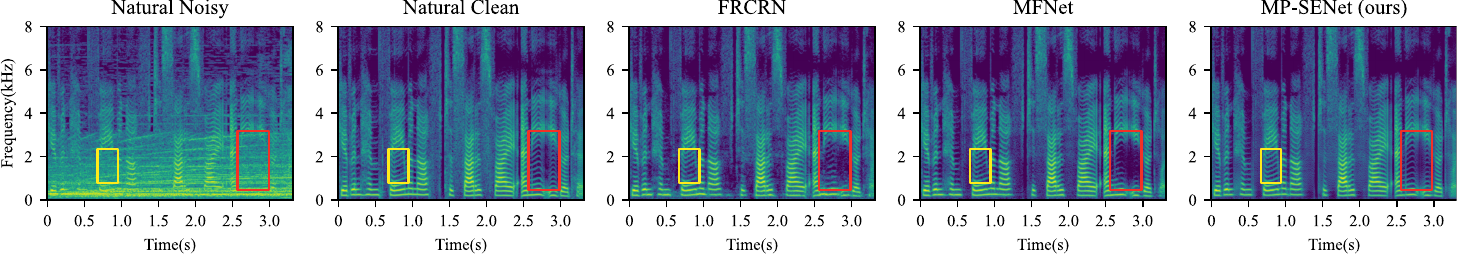}
\label{fig: specs_denoising_dns}}
\caption{Spectrogram visualization of the noisy speech, clean speech, and speech waveforms denoised by SOTA baseline methods and our proposed MP-SENet.}
\label{fig: specs_denoising}
\end{figure*}

\section{Results and Analysis}
\label{sec: V}
\subsection{Speech Denoising}
\subsubsection{Evaluation metrics}
For the evaluation on the VoiceBank+DEMAND dataset, five commonly used objective evaluation metrics were chosen to evaluate the enhanced speech quality, including wideband PESQ (WB-PESQ), STOI, and three composite measures \cite{hu2007evaluation} (i.e., CSIG, CBAK, and COVL).
WB-PESQ was used to evaluate the wideband speech perceptual quality, which ranges from -0.5 to 4.5.
STOI was used to measure speech intelligibility, which ranges from 0 to 1.
CSIG, CBAK, and COVL measure signal distortion, background noise intrusiveness, and overall effect, respectively. 
All these three mean opinion score (MOS) based metrics range from 1 to 5.
For the evaluation on the DNS Challenge dataset, besides the WB-PESQ and the STOI, we further adopted the narrowband PESQ (NB-PESQ) to evaluate the narrowband speech perceptual quality and the scale-invariant signal-to-distortion ratio (SI-SDR) \cite{le2019sdr} to quantify the distortion between the enhanced and clean speech signals.
For all the metrics, higher values indicate better performance.

\subsubsection{Comparison with baseline methods}
We evaluated the performance of our proposed MP-SENet separately on the VoiceBank+DEMAND dataset and the large-scale DNS Challenge dataset, with different baseline models used for each of these datasets.
\begin{itemize}[left=0em]
\item {}{\textbf{Evaluation on the VoiceBank+DEMAND Dataset}}: 
we compared our proposed MP-SENet with several representative time-domain and TF-domain methods. In the time-domain category, we included SEGAN \cite{pascual2017segan}, DEMUCS \cite{defossez2020real}, and SE-Conformer \cite{kim2021se}. 
For the TF-domain methods, we selected MetricGAN \cite{fu2019metricgan}, MetricGAN+ \cite{fu2021metricgan+}, PHASEN \cite{yin2020phasen}, and four SOTA methods, namely DPT-FSNet \cite{dang2022dpt}, TridentSE \cite{yin2022tridentse}, DB-AIAT \cite{yu2022dual}, and CMGAN \cite{abdulatif2022cmgan}.
As depicted in Table~\ref{tab: denoising_vb}, our proposed MP-SENet outperformed all other speech denoising methods across all metrics, demonstrating its strong denoising capability.
It was noteworthy that the time-domain methods generally performed inferior to TF-domain methods, indicating the importance of TF-domain characteristics for the speech denoising task.
Within the TF-domain methods, while our proposed MP-SENet and PHASEN both used magnitude and phase spectra as input conditions, MP-SENet exhibited significant improvements of 0.61, 0.60, 0.44, and 0.72 in WB-PESQ, CSIG, CBAK, and COVL scores with a nine times smaller model size compared to PHASEN.
Compared to the other four TF-domain SOTA approaches, our proposed MP-SENet with explicit phase optimization still excelled among these metrics.
Especially, the higher WB-PESQ score signified better magnitude spectra generation, indicating that our proposed MP-SENet with explicit phase optimization can further alleviate the compensation effect between magnitude and phase, thereby enhancing the speech perceptual quality.

For more evidence, we visualized the spectrograms of speech waveforms enhanced by DB-AIAT, CMGAN, and our proposed MP-SENet as shown in Fig.~\ref{fig: specs_denoising_vb}.
By comparing the contents of boxes with the same color, it is evident that our proposed MP-SENet achieved better denoising performance compared to DB-AIAT and CMGAN, and effectively preserved the harmonic structures of the speech. 
This further confirmed the effectiveness of explicit phase optimization in promoting the generation of high-fidelity magnitude spectra.

\begin{table}[t!]\Huge
  \caption{Experimental results for speech denoising methods evaluated on the DNS Challenge evaluation set $w/o$ reverberation.}
  \label{tab: denoising_dns}
  \centering
  \resizebox{\linewidth}{!}{
  \begin{tabular}{lcccccc}
    \toprule
    \multirow{2}{*}{Method} & \multirow{2}{*}{Year} & \multirow{2}{*}{\#Param.} & \multirow{2}{*}{WB-PESQ} & \multirow{2}{*}{NB-PESQ} & STOI & SI-SDR \\
    & & & & & (\%) & (dB)\\
    \midrule
    Noisy                              & -    & -     & 1.58 & 2.45 & 91.52 &  9.07 \\
    \midrule
    FullSubNet\cite{hao2021fullsubnet} & 2021 & 5.64M & 2.78 & 3.31 & 96.11 & 17.29 \\
    CTSNet\cite{li2021two}             & 2021 & 4.35M & 2.94 & 3.42 & 96.21 & 16.69 \\
    TaylorSENet\cite{li2022taylor}     & 2022 & 5.40M & 3.22 & 3.59 & 97.36 & 19.15 \\
    FRCRN\cite{zhao2022frcrn}          & 2022 & 6.90M & 3.23 & 3.60 & 97.69 & 19.78 \\
    MFNet\cite{liu2023mask}            & 2023 & -     & 3.43 & 3.74 & 97.78 & 20.31 \\
    \textbf{MP-SENet}                  & 2024 & 2.26M & \textbf{3.62} & \textbf{3.92} & \textbf{98.16} & \textbf{21.03} \\
    \bottomrule
  \end{tabular}}
\end{table}

\item {}{\textbf{Evaluation on the DNS Challenge Dataset}}: 
We further compared our proposed MP-SENet with several TF-domain speech denoising methods on the DNS Challenge dataset, including two complex spectral masking-based methods (i.e., FullSubNet \cite{hao2021fullsubnet} and FRCRN \cite{zhao2022frcrn}), two decoupling-style methods (i.e., CTSNet \cite{li2021two} and TaylorSENet \cite{li2022taylor}), and a short-time discrete cosine transform (STDCT) spectrum-based method (i.e., MFNet).
As depicted in Table~\ref{tab: denoising_dns}, our proposed MP-SENet significantly outperformed these baseline methods in terms of all the metrics with the smallest parameter count.
Especially, our MP-SENet demonstrated huge advantages in PESQ metrics, which reflected the improvement in speech perceptual quality.
Compared to the STFT spectrum-based methods, our proposed MP-SENet achieved superior performance due to the explicit estimation of the phase.
Although MFNet achieved implicit magnitude-phase recovery by predicting a real-valued STDCT spectrum, its performance in waveform generation tasks was still inferior to that of the STFT spectrum, which may be attributed to the fact that an over-complete Fourier basis contributed to improved training stability \cite{gritsenko2020spectral}.

We also visualized the spectrograms of speech waveforms enhanced by our proposed MP-SENet and the other two SOTA baselines, including FRCRN and MFNet.
As shown in Fig.~\ref{fig: specs_denoising_dns}, we can observe that although the noise distorted the fundamental frequency and the low-order harmonics in the noisy speech, both the baseline models and our MP-SENet can effectively restore them, which could be because the energy of them was relatively high, making them easier to be separated from the noisy components.
However, as shown in the yellow and red boxes in the figure, when the higher-order harmonics were disrupted, the baseline models, lacking explicit phase optimization, failed to adequately restore the harmonic structures during enhancement.
This further confirmed the importance of accurate phase estimation for the speech denoising task.

\end{itemize}

\begin{figure*}[t!]
  \centering
  \includegraphics[width=\textwidth]{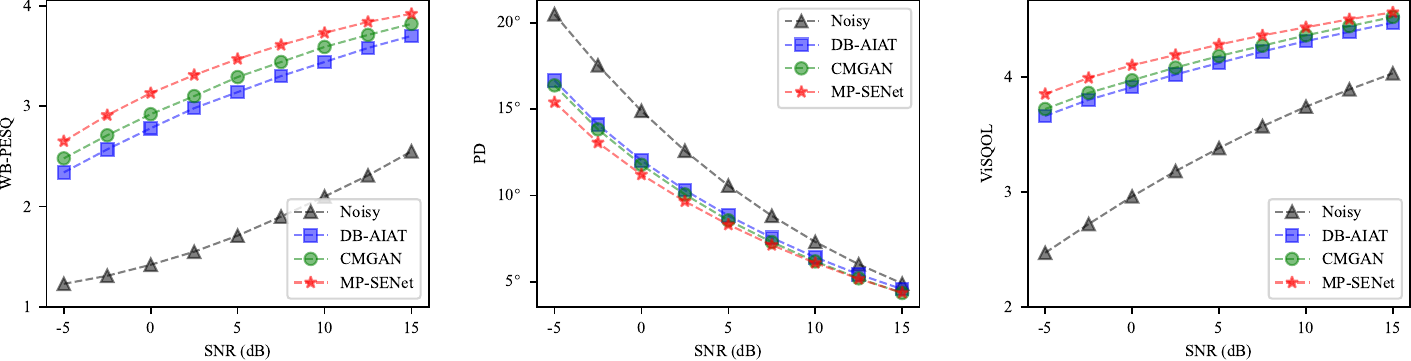}
  \caption{WB-PESQ, PD, and ViSQOL metrics of the noisy speech and speech waveforms enhanced by DB-AIAT, CMGAN, and our proposed MP-SENet on the re-noised Voice Bank test set with varying SNR conditions.}
  \label{fig: specs_metric}
\end{figure*}

\subsubsection{SNR-wise Analysis Experiment}
Paliwal \emph{et al.} \cite{paliwal2011importance} emphasized the importance of phase information for speech denoising tasks in low-SNR circumstances.
Theoretically, The noisy phase $\bm{Y}_p$ can be derived as follows \cite{zheng2018phase}:
\begin{equation}
	\begin{aligned}
	   \bm{Y}_p &= \mathrm{Angle}(\bm{X}_me^{j\bm{X}_p} + \bm{N}_me^{j\bm{N}_p}) \\
                &= \bm{X}_p + \mathrm{Angle}\bigg(1 + \frac{\bm{N}_m}{\bm{X}_m} e^{j(\bm{N}_p-\bm{X}_p)}\bigg),
    \end{aligned}
    \label{phase in low-SNR}
\end{equation}
where $\bm{N}_m \in \mathbb{R}^{T\times F}$ and $\bm{N}_p \in \mathbb{R}^{T\times F}$ denote the magnitude and phase spectra of the noise signal $\bm{n} \in \mathbb{R}^{L}$, respectively. $\mathrm{Angle}(\cdot)$ calculates the phase spectrum of a complex spectrum.
When in a high-SNR circumstance, the clean magnitude spectrum $\bm{X}_m$ is much larger than the noise magnitude spectrum $\bm{N}_m$, the second term in Eq.~\ref{phase in low-SNR} would approach zero, indicating that the clean phase spectrum $\bm{X}_p$ can be approximated by the noisy phase spectrum $\bm{Y}_p$. 
So we hypothesized that as the SNR increases, the contribution of phase information to the performance of the SE model would diminish.

To experimentally verify our hypothesis, we conducted an analysis experiment in which we tested the performance of our proposed MP-SENet under different SNR circumstances.
Additionally, we selected DB-AIAT and CMGAN, which implicitly enhance the phase, as comparative systems to validate the importance of explicit phase enhancement.
With the clean test set of the Voice Bank corpus, we re-noised it with five types of unseen noise (i.e., living room, office space, bus, open area cafeteria, and a public square) under separate SNR conditions, which range from -5 dB to 15 dB with intervals of 2.5 dB.
Subsequently, we employed these test sets with varying SNRs to evaluate the performance of DB-AIAT, CMGAN, and our proposed MP-SENet.
To quantitatively measure the phase estimation, we used the phase distance (PD) metric \cite{choi2018phase} which was defined as the average angle difference weighted by target magnitude between the enhanced speech and the target speech in the TF domain.
The PD can be formulated using our proposed anti-wrapping function as follows:
\begin{equation}
    PD = \frac{360^\circ}{\pi}\sum_{t,f}\frac{\bm{X}_m[t,f]}{\sum_{t',f'}\bm{X}_m[t',f']}f_{AW}(\bm{X}_p[t,f] - \bm{\hat{X}}_p'[t,f]),
\end{equation}
where $\bm{\hat{X}}_p'$ denotes the re-extracted phase spectrum from the enhanced speech $\bm{\hat{x}}$.
The PD metric ranges from $0^{\circ}$ to $180^{\circ}$, the lower the better.
Furthermore, since WB-PESQ is a magnitude-related metric that can not reflect the overall speech quality, we adopted the virtual speech quality objective listener (ViSQOL) \cite{chinen2020visqol}, which utilizes a spectral-temporal measure of similarity between reference and test speech signals to produce a mean opinion score - listening quality objective (MOS-LQO) score. 
The ViSQOL metric ranges from 1 to 5, the higher the better.

The experimental results were plotted into curves as illustrated in Fig.~\ref{fig: specs_metric}.
Firstly, it is obvious that our proposed MP-SENet outperformed the other two baselines among the three metrics, demonstrating its superiority in refining both magnitude and phase and improving overall speech quality.
For the WB-PESQ metric, both the baseline models and MP-SENet exhibited relatively stable improvements compared to the noisy speech across different SNRs.
For the PD metric, the enhanced speech showed a significant reduction in phase distance compared to the noisy speech at very low SNRs. 
However, as the SNR increased, the gap in the PD metric between the enhanced speech and the noisy speech diminished and eventually tended toward zero.
At the same time, the difference in the ViSQOL metric between them was also narrowing, which aligned with our hypothesis and highlighted the significance of recovering phase information in low-SNR circumstances.
It is noteworthy that, the two baseline models with implicit phase optimization, almost overlapped in PD and ViSQOL.
However, our MP-SENet showed apparent improvements in these two metrics in low-SNR conditions, which confirmed the contribution of explicit phase estimation to the overall enhanced speech quality.

\subsubsection{Ablation Studies}
\label{sec: ablation}
To investigate the role of each key component in the MP-SENet, we took the speech denoising task as the ablation example and performed ablation studies on the VoiceBank+DEMAND dataset.
We used the WB-PESQ, PD, SI-SDR, and ViSQOL metrics to evaluate the magnitude restoration, phase restoration, distortion of the enhanced waveform, and overall speech quality, respectively.
The ablation results are presented in Table.~\ref{tab: ablation}.

\begin{figure*}[t!]
  \centering
  \includegraphics[width=\textwidth]{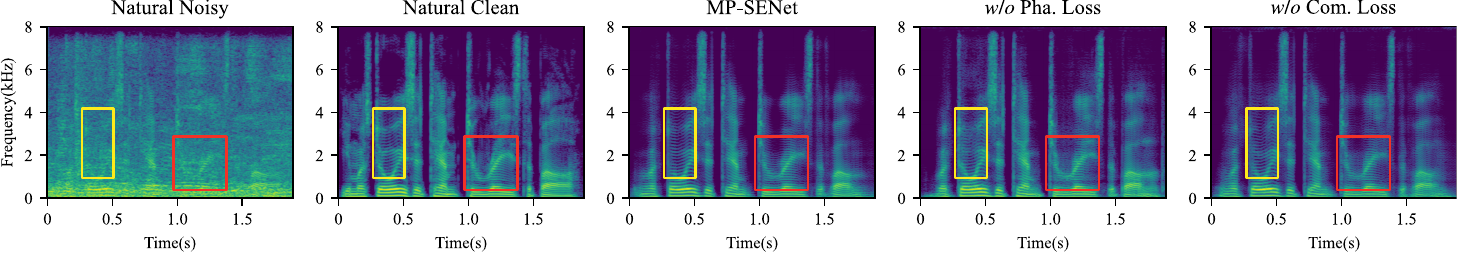}
  \caption{Spectrogram visualization of the noisy speech, clean speech, and speech waveforms enhanced our proposed MP-SENet with different combinations of phase optimization approaches on the VoiceBank+DEMAND dataset.}
  \label{fig: specs_loss}
\end{figure*}
\begin{table}[t!]\Huge
  \caption{Experimental results of the ablation studies on the VoiceBank+DEMAND dataset.}
  \label{tab: ablation}
  \centering
  \resizebox{\linewidth}{!}{
  \begin{tabular}{lccccc}
    \toprule
    Method                 & \#Param.       & WB-PESQ       & PD ($^\circ$)  & SI-SDR (dB)    & ViSQOL \\
    \midrule
    MP-SENet               & 2.26M          & \textbf{3.60} & 7.28           & 19.47          & \textbf{4.34} \\
    \midrule
    \multicolumn{6}{c}{\textbf{Ablation on Model Structure}} \\
    \midrule
    $w/$ Conformer        & \textbf{1.71M} & 3.56           & 7.35          & 19.38          & 4.30 \\
    Magnitude Only         & 1.99M          & 3.44           & 8.00          & 18.37          & 4.28 \\
    Complex Only           & 2.26M          & 3.58           & 7.56          & 18.57          & 4.28 \\
    \midrule
    \multicolumn{6}{c}{\textbf{Ablation on Loss Functions}} \\
    \midrule
    $w/o$ Pha. Loss 	   & 2.26M          & 3.52           & 7.65          & 19.25          & 4.32 \\
    $w/o$ Com. Loss 	   & 2.26M          & 3.57           & 7.48          & 18.60          & 4.33 \\
    $w/o$ Con. Loss 	   & 2.26M          & 3.58           & \textbf{7.26} & 18.96          & 4.33 \\
    $w/o$ Metric Disc.     & 2.26M          & 3.44           & 7.38          & \textbf{20.05} & 4.25 \\
    \bottomrule
  \end{tabular}}
\end{table}

We first conducted ablation studies on the model structure.
To compare the modeling capability between the Transformers and the Conformers used in CMGAN, we replaced the Transformers in the TF-Transformer blocks with Conformers (denoted as ``$w/$ Conformer'').
As a result, the ablation of the Transformers resulted in performance degradation, which may be attributed to the fact that Transformers have better modeling capabilities than Conformers.
However, it was noteworthy that our MP-SENet with Conformers still achieved a PESQ of 3.56 with a smaller model size of 1.71M, which significantly outperformed those of CMGAN in Fig.~\ref{tab: denoising_vb} (PESQ of 3.41 and model size of 1.83M).
This parallel comparison highlighted the contribution of our proposed magnitude-phase parallel prediction architecture and explicit phase optimization.
Subsequently, we investigated the effect of magnitude-phase parallel modeling.
We first ablated the phase decoder to only conduct magnitude enhancement (denoted as ``Magnitude Only''), in which the enhanced magnitude was combined with the noisy phase to reconstruct the enhanced speech waveform.
As a result, all the metrics significantly degraded, which would be attributed to that without explicit phase prediction, the magnitude would significantly compensate for the noisy phase, resulting in both poor magnitude restoration and phase accuracy. 
We then re-designed the magnitude mask decoder and the phase decoder to the real-part decoder and imaginary-part decoder, which used complex spectral mapping to directly output the real and imaginary parts of the short-time complex spectrum (denoted as ``Complex Only'').
Consequently, all the metrics marginally degraded, which demonstrated the superiority of explicit magnitude and phase modeling, and also indicated that complex spectrum-based methods can still achieve exceptional enhancement performance under the guidance of explicit magnitude and phase optimization.
Overall, the above experiments demonstrate the effectiveness of the explicit magnitude-phase modeling we proposed.

Furthermore, we conducted ablation studies on the loss functions.
To investigate the effects of phase optimization approaches, we conducted ablation studies on the phase spectrum loss (denoted as ``$w/o$ Pha. loss") and complex spectrum loss (denoted as ``$w/o$ Com. loss"), which explicitly and implicitly optimized the phase, respectively.
Results demonstrated that both of them contributed to the overall performance, but explicit phase optimization played a more pivotal role.
We further visualized the spectrograms of the speech waveforms enhanced by these two ablation models and our proposed MP-SENet as illustrated in Fig~\ref{fig: specs_loss}.
It can be clearly observed that after ablating the phase loss, the harmonic structures were significantly distorted, while this impact was slighter when ablating the complex spectrum loss. 
This further highlighted the importance of explicit phase optimization in alleviating the compensation effect and restoring the harmonic structures.
Subsequently, we solely ablated the STFT consistency loss (denoted as $w/o$ Con. loss) and the results demonstrated that the STFT consistency did contribute to the model performance.
Finally, we removed the metric discriminator (denoted as ``$w/o$ Metric disc.") to assess the effect of the human-perception-related metric loss.
Ablation results demonstrated that introducing the metric loss significantly enhanced PESQ and ViSQOL but marginally sacrificed SI-SDR. 
This suggested that the slight degradation in SI-SDR did not affect the overall enhanced speech quality.

\subsection{Speech Dereverberation}
\subsubsection{Evaluation metrics}
For the speech dereverberation task, we used the speech quality metrics provided by the REVERB Challenge, including WB-PESQ, cepstral distance (CD), log-likelihood ratio (LLR), frequency-weighted segmental signal-to-noise ratio (FWSegSNR), and signal-to-reverberation modulation energy ratio (SRMR).
Since the `RealData' of the REVERB Challenge evaluation set lacked corresponding clean reference speech, only the SRMR metric was employed to evaluate the model's dereverberation performance in real scenarios.
Lower CD and LLR, as well as higher FWSegSNR and SRMR, indicate better performance.

\begin{figure*}[t!]
  \centering
  \includegraphics[width=\textwidth]{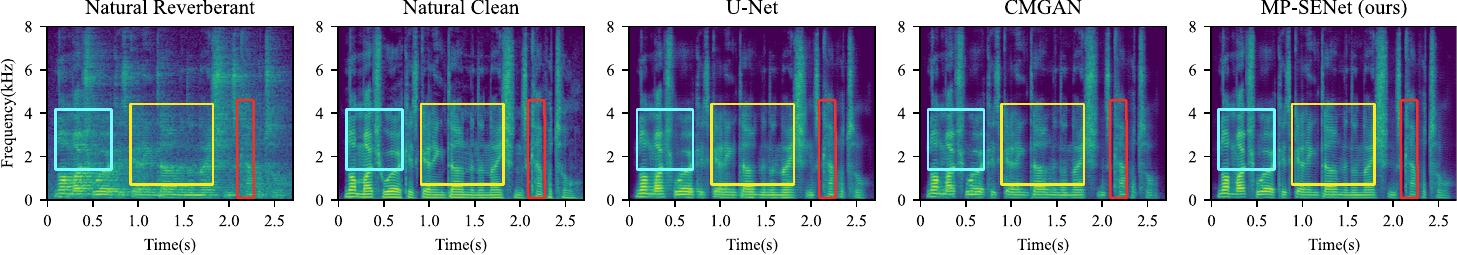}
  \caption{Spectrogram visualization of the reverberant speech, clean speech, and speech waveforms dereverberated by U-Net, CMGAN, and our proposed MP-SENet on the REVERB Challenge `SimData' set.}
  \label{fig: specs_dereverberation}
\end{figure*}
\begin{figure*}[t!]
  \centering
  \includegraphics[width=\textwidth]{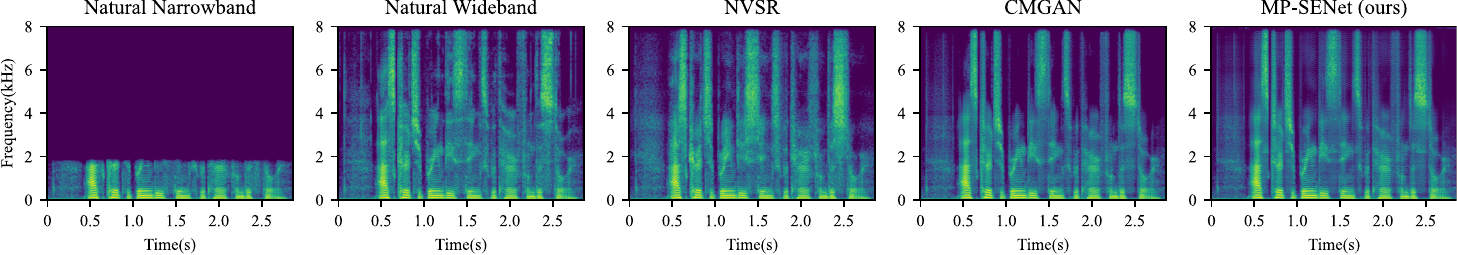}
  \caption{Spectrogram visualization of the narrowband speech, wideband speech, and speech waveforms extended by NVSR, CMGAN, and our proposed MP-SENet on the VCTK dataset with the source sampling rate of 4 kHz and target sampling rate of 16 kHz.}
  \label{fig: specs_bwe}
\end{figure*}

\subsubsection{Comparison with baseline methods}
We compared our proposed MP-SENet with a signal processing-based method, i.e., WPE \cite{nakatani2010speech}, as well as four DNN-based methods, including the magnitude mapping-based UNet \cite{ernst2018speech}, and three complex spectrum-based methods (i.e., CMGAN \cite{abdulatif2022cmgan}, SkipConvGAN \cite{kothapally2022skipconvgan}, and DCN \cite{kothapally2024monaural}).
The objective evaluation results are presented in Table.~\ref{tab: dereverberation}.
It's obvious that DNN-based methods exhibited more robust dereverberation capabilities compared to the signal processing-based WPE.
Among the DNN-based methods, UNet which only enhanced the magnitude while retaining the reverberant phase apparently lagged behind other phase-aware methods. 
Compared to the complex spectrum-based methods, our proposed MP-SENet achieved optimal results in most metrics on the `SimData', with suboptimal results for the remaining LLR and SRMR metrics.
However, the MP-SENet achieved the optimal SRMR on the `RealData', indicating that it exhibited better robustness in real reverberant environments.
\begin{table}[t!]\Huge
  \caption{Experimental results for speech dereverberation methods evaluated on the REVERB Challenge evaluation set.}
  \label{tab: dereverberation}
  \centering
  \resizebox{\linewidth}{!}{
  \begin{tabular}{lc|ccccc|c}
    \toprule
    \multirow{2}{*}{Method} & \multirow{2}{*}{Year} & \multicolumn{5}{c|}{SimData} & RealData \\
    \cmidrule{3-8}
    & & PESQ & CD & LLR & SNR$_{fw}$ & SRMR & SRMR \\
    \midrule
    Reverberant & - & 1.50 & 3.97 & 0.58 & 3.62 & 3.69 & 3.18 \\
    \midrule
    WPE\cite{nakatani2010speech} & 2010 & 1.72 & 3.75 & 0.51 & 4.90 & 4.22 & 3.98 \\
    UNet\cite{ernst2018speech} & 2018 & - & 2.50 & 0.40 & 10.70 & 4.88 & 5.58 \\ 
    SkipConvGAN \cite{kothapally2022skipconvgan} & 2022 & 2.91 & 2.32 & \textbf{0.23} & 11.90 & \textbf{5.89} & 6.36 \\
    CMGAN \cite{abdulatif2022cmgan} & 2023 & - & 2.25 & 0.31 & 11.74 & 5.47 & 6.55 \\
    DCN \cite{kothapally2024monaural} & 2024 & 2.94 & 2.00 & \textbf{0.23} & 13.33 & 5.27 & 6.48 \\
    \textbf{MP-SENet} & 2024 & \textbf{2.97} & \textbf{1.97} & 0.24 & \textbf{14.07} & 5.51 & \textbf{6.67} \\
    \bottomrule
  \end{tabular}}
\end{table}
To intuitively compare the dereverberation performance of different methods, we also visualized the spectrograms of speech waveforms dereverberated by UNet, CMGAN, and our proposed MP-SENet as shown in Fig.~\ref{fig: specs_dereverberation}.
By comparing the contents of boxes with the same color, We can observe that UNet, due to its utilization of the reverberant phase, inevitably suffered from magnitude distortion during iSTFT.
Although CMGAN partially restored the phase information through implicit phase optimization, the imprecise phase estimation still induced the compensation effect, resulting in poor restoration of the harmonic structures.
Nevertheless, our proposed MP-SENet, utilizing explicit phase modeling and optimization, effectively restored continuous and intact harmonic structures from the smeared spectra, further improving the quality of the enhanced speech.

\begin{table}[t!]\Huge
  \caption{Experimental results for speech bandwidth extension methods on the VCTK dataset. ``*'' denotes the reproduced results.}
  \label{tab: bwe}
  \centering
  \resizebox{\linewidth}{!}{
  \begin{tabular}{lc|ccc|ccc}
    \toprule
    \multirow{2}{*}{Methods} & \multirow{2}{*}{Year} & \multicolumn{3}{c|}{8 kHz $\rightarrow$ 16 kHz} & \multicolumn{3}{c}{4 kHz $\rightarrow$ 16 kHz} \\
    \cmidrule{3-8}
    & & WB-PESQ & LSD & ViSQOL & WB-PESQ & LSD & ViSQOL \\
    \midrule
    AudioUNet\cite{kuleshov2017audio} & 2017 & 3.68 & 1.32 & - & 3.39 & 1.40 & - \\
    TFNet\cite{lim2018time} & 2018 & 3.72 & 0.99 & - & 3.48 & 1.22 & - \\
    AECNN\cite{wang2021towards} & 2021 & 3.91 & 0.88 & - & 3.64 & 0.95 & - \\
    NVSR\cite{liu2022neural} & 2022 & 3.56 & 0.79 & 4.52 & 2.40 & 0.95 & 4.11 \\
    CMGAN$^*$\cite{abdulatif2022cmgan} & 2023 & 4.24 & 0.81 & 4.70 & 3.73 & 1.04 & 4.35 \\
    \textbf{MP-SENet} & 2024 & \textbf{4.28} & \textbf{0.66} & \textbf{4.72} & \textbf{3.78} & \textbf{0.81} & \textbf{4.42} \\
    \bottomrule
  \end{tabular}}
\end{table}

\subsection{Speech Bandwidth Extension}
\subsubsection{Evaluation metrics}
For the speech BWE task, we first adopted the WB-PESQ and ViSQOL to evaluate the magnitude-related speech perceptual quality and the overall speech quality, respectively.
Additionally, log-spectral distance (LSD) \cite{gray1976distance} was utilized to measure the logarithmic distance between two magnitude spectra in dB.
lower LSD values indicate better performance.

\subsubsection{Comparison with baseline methods}
We compared MP-SENet with two time-domain methods (i.e., AudioUNet \cite{kuleshov2017audio} and AECNN \cite{wang2021towards}), two TF-domain methods (i.e., NVSR\cite{liu2022neural} and CMGAN \cite{abdulatif2022cmgan}), and a hybrid-domain method (i.e., TFNet \cite{lim2018time}).
The experimental results are presented in Table.~\ref{tab: bwe}.
It is evident that our proposed MP-SENet achieved the optimal performance with the source sampling rates of both 4 kHz and 8 kHz.
Overall, time-domain methods still appeared to be slightly inferior to TF-domain methods, highlighting the importance of low-frequency TF-domain features for recovering high-frequency components.
Within the TF-domain methods, we directly visualized the spectrograms of speech waveforms extended by NVSR, CMGAN, and our proposed MP-SENet with a subsampling rate of 4.
Combining the results from Table.~\ref{tab: bwe} and the spectrograms illustrated in Fig.~\ref{fig: specs_bwe}, we can observe that NVSR restored wideband waveforms from the extended mel-spectrogram using a neural vocoder, resulting in higher energy in the high-frequency part compared to real wideband speech, thus leading to the collapse of the PESQ metric. 
Additionally, since it did not utilize the low-frequency phase information, its performance in recovering high-frequency harmonics was relatively poorer.
Compared to CMGAN, our proposed MP-SENet exhibited a significant improvement in LSD. 
By comparing the high-frequency components of their spectrograms, we can observe that the MP-SENet demonstrated a strong capability to restore high-frequency harmonics, which confirms the importance of phase information in speech BWE tasks.
It is worth noting that MP-SENet adopted the same model structure as the previous speech denoising and dereverberation tasks for the speech BWE task, which achieved an effective extension of narrowband magnitude spectrum through a high-frequency mask, demonstrating the versatility of our proposed parallel magnitude-phase enhancement architecture.

\vspace{-1mm}
\section{Conclusion}
\label{sec: VI}
In this paper, we proposed a TF-domain monaural SE model, named MP-SENet, which explicitly enhanced both magnitude and phase spectra in parallel.
The MP-SENet comprised an encoder-decoder architecture, where the encoder encoded distorted magnitude and phase spectra to TF-domain representations, and the parallel magnitude mask decoder and phase decoder output the enhanced magnitude and phase spectra, respectively.
The encoder and decoder are bridged by TF-Transformers to capture time and frequency dependencies. 
The major contribution of the MP-SENet lay in the explicit modeling and optimization of the phase spectrum, which provided a new solution for the current SE framework, and further elevated the SE performance to a new level.
Experimental results demonstrated that our proposed MP-SENet achieved SOTA SE performance across speech denoising, dereverberation, and bandwidth extension tasks with a unified framework.
Spectrogram visualizations demonstrated the powerful harmonic restoration capability of our MP-SENet, indicating that explicit phase estimation can effectively alleviate the compensation effect.
Further reducing the model parameters and improving the inference efficiency will be the focus of our future work.
\vspace{-1mm}

\bibliographystyle{IEEEtran}
\bibliography{mybib}

\end{document}